# *Replicating Physical Motion with Minkowskian Isorefractive Spacetime Crystals*

**Filipa R. Prudêncio[1,2*], and Mário G. Silveirinha[1]**

[1]University of Lisbon – Instituto Superior Técnico and Instituto de Telecomunicações, Avenida Rovisco Pais 1, 1049-001 Lisbon, Portugal

[2]Instituto Universitário de Lisboa (ISCTE-IUL), Avenida das Forças Armadas 376, 1600-077 Lisbon, Portugal

## Abstract

Here, we show that isorefractive spacetime crystals with a travelling-wave modulation may mimic rigorously the response of moving material systems. While generic spacetime crystals are characterized by a bi-anisotropic coupling in the co-moving frame, isorefractive crystals have a response that is observer independent, which leads to isotropic constitutive relations free of bianisotropy. We show how to take advantage of this property in the calculation of the band diagrams of isorefractive spacetime crystals in the laboratory frame and in the study of the synthetic Fresnel drag. Furthermore, we discuss the impact of considering either a Galilean or a Lorentz transformation in the homogenization of spacetime crystals, showing that the effective response is independent of the considered transformation.

[*] Corresponding author: filipa.prudencio@lx.it.pt

# I. Introduction

In recent years, time-varying material responses have opened up many interesting opportunities in metamaterials and in other light-based platforms [1-15]. Time modulated materials may be useful to design magnetless non-reciprocal systems, such as unidirectional guides and isolators [4, 5, 6, 9]. The wave phenomena in time-modulated systems can be quite rich and peculiar [1-15].

A particular class of spacetime crystals has attracted considerable attention due to the relative simplicity of modeling and system design: the "travelling-wave" spacetime crystals [7-8, 16-22]. In a travelling-wave crystal the material parameters, let us say the permittivity, depends on space and time as $\varepsilon(\mathbf{r},t) = \varepsilon(\mathbf{r} - \mathbf{v}t)$, where $\mathbf{r} = (x, y, z)$ is a generic point of space and $\mathbf{v}$ is the modulation speed. Remarkably, the travelling-wave modulation of the material parameters may induce a synthetic Fresnel drag in the long wavelength regime, such that waves propagating in the spacetime crystal are dragged towards a preferred direction of space, either parallel or anti-parallel to $\mathbf{v}$ [19]. Interestingly, these effects may be conveniently described using homogenization theory [18, 20, 21]. Specifically, a travelling-wave crystal formed by layered dielectrics behaves effectively as a bianisotropic nonreciprocal material in the long wavelength limit. In the simplest case, where the crystal is formed by simple isotropic materials, the magneto-electric coupling tensor is anti-symmetric, corresponding to a standard "moving-medium" coupling [19, 20, 23]. Curiously, by controlling the optical axes and the anisotropy of the materials that form the spacetime crystal, it is possible to engineer nearly arbitrary (Hermitian) nonreciprocal couplings [21]. For example, it was shown in Ref. [21] that anisotropic spacetime crystals with a suitable glide-rotation symmetry may exhibit an isotropic effective Tellegen (axion) response in the long wavelength limit. Different from the moving medium coupling, the Tellegen response is determined by a "scalar", which is the simplest example of a symmetric magneto-electric tensor [24-25]. Tellegen materials are nonreciprocal and thereby can be potentially useful to realize unidirectional devices [26-27] and systems with nontrivial topological properties [28-29].

While previous works have shown that crystals with a travelling-wave modulation can effectively mimic physical motion [18-21], the analogy is imperfect in many ways and typically only holds true in the long wavelength limit. For example, the velocity of equivalent moving medium $\mathbf{v}_\mathrm{D}$ typically does not match the modulation speed of the crystal $\mathbf{v}$, and most puzzling the sign of the two velocities can be different [19]. In fact, a spacetime modulated dielectric crystal is not equivalent to a moving dielectric crystal. The reason will be developed in detail in the following sections, but essentially boils down to the fact that in the co-moving frame –where the material response is time-independent– the response of a dielectric crystal is by definition free of magneto-electric coupling, while the response of a spacetime crystal is bianisotropic.

Notwithstanding with the constraints discussed in the previous paragraph, here we show that there is a particular subclass of spacetime crystals that may replicate *exactly* the response of moving material bodies for *any* frequency of operation. Specifically, we show that isorefractive spacetime crystals – formed by materials with a constant refractive index, $n = \sqrt{\varepsilon_\mathrm{r}\mu_\mathrm{r}} = const.$ – have an electromagnetic response identical to that of a moving dielectric crystal in a suitable "vacuum" background. In particular, our analysis unveils that isorefractive crystals may be the ideal platforms to mimic physical motion using spacetime modulations. Furthermore, we highlight how by using generalized Lorentz transformations it is possible to determine in a rather straightforward manner the dispersion properties of spacetime modulated isorefractive crystals and characterize the synthetic Fresnel drag in the long wavelength limit. Finally, we discuss the impact of using relativistic (Lorentz) and non-relativistic (Galilean) transformations in the analysis of spacetime crystals. It is demonstrated that while both types of transformation predict the same physics and the same effective response in the long wavelength limit, the use of Lorentz transformations greatly simplifies the analysis in the case of isorefractive crystals.

## II. Isorefractive Spacetime Crystals

*A. Coordinate transformations and constitutive relations of a spacetime crystal*

We are interested in spacetime crystals with a travelling-wave modulation characterized by isotropic constitutive relations,

$$\mathbf{D}(\mathbf{r},t) = \varepsilon_0 \varepsilon(\mathbf{r} - \mathbf{v}t) \mathbf{E}(\mathbf{r},t), \quad \mathbf{B}(\mathbf{r},t) = \mu_0 \mu(\mathbf{r} - \mathbf{v}t) \mathbf{H}(\mathbf{r},t). \quad (1)$$

We shall assume without loss of generality that the modulation speed is along the *x*-direction, $\mathbf{v} = v\hat{\mathbf{x}}$, so that $\varepsilon(\mathbf{r} - \mathbf{v}t)$ stands for a function of the type $\varepsilon(x - vt, y, z)$, determined by only three independent degrees of freedom.

Evidently, the material response is time-independent in an inertial (co-moving) frame that moves with speed $\mathbf{v}$ with respect to the laboratory frame. In this article, we link the coordinates of the two frames through a generalized Lorentz transformation of the type:

$$x' = \gamma(x - vt), \quad y' = y, \quad z' = z \quad (2a)$$

$$t' = \gamma\left(t - x\frac{v}{c_0^2}\right), \quad (2b)$$

with $\gamma = (1 - v^2/c_0^2)^{-1/2}$ the Lorentz factor. Here, $c_0$ is a free positive parameter with unities of velocity. If $c_0$ is taken identical to the speed of light in vacuum, $c$, we recover the standard Lorentz transformation, whereas if $c_0 = \infty$ we get a simple Galilean coordinate transformation.

The structure of the Maxwell's equations is preserved by any generalized Lorentz transformation, provided the electromagnetic fields are transformed in the usual manner [23, 30]:

$$\mathbf{E}'_\| = \mathbf{E}_\|, \qquad\qquad \mathbf{B}'_\| = \mathbf{B}_\|, \qquad\qquad (3a)$$

$$\mathbf{E}'_\perp = \gamma(\mathbf{E}_\perp + \mathbf{v}\times\mathbf{B}), \qquad\qquad \mathbf{B}'_\perp = \gamma\left(\mathbf{B}_\perp - \frac{1}{c_0^2}\mathbf{v}\times\mathbf{E}\right), \qquad\qquad (3b)$$

and

$$\mathbf{D}'_\| = \mathbf{D}_\|, \qquad\qquad \mathbf{H}'_\| = \mathbf{H}_\|, \qquad\qquad (4a)$$

$$\mathbf{D}'_\perp = \gamma\left(\mathbf{D}_\perp + \frac{1}{c_0^2}\mathbf{v}\times\mathbf{H}\right), \qquad\qquad \mathbf{H}'_\perp = \gamma(\mathbf{H}_\perp - \mathbf{v}\times\mathbf{D}). \qquad\qquad (4b)$$

Here $\|$ and $\perp$ represent the field components parallel and perpendicular to the velocity. Evidently, the primed fields have a strong physical meaning when $c_0 = c$, as in that case they coincide with the physical fields evaluated in the relevant inertial frame. When $c_0 \neq c$, the primed fields should be simply regarded as auxiliary fields that are introduced to simplify the mathematical treatment of the wave propagation problem in the spacetime crystal.

As shown in previous works [20, 21], the generalized Lorentz transformation leads to the following constitutive relations for the primed fields (compare with Eq. (1)):

$$\begin{pmatrix}\mathbf{D}'\\ \mathbf{B}'\end{pmatrix} = \begin{pmatrix}\varepsilon_0\overline{\varepsilon}' & \frac{1}{c}\overline{\xi}'\\ \frac{1}{c}\overline{\zeta}' & \mu_0\overline{\mu}'\end{pmatrix}\begin{pmatrix}\mathbf{E}'\\ \mathbf{H}'\end{pmatrix} \qquad\qquad (5a)$$

where the transformed permittivity, permeability and magneto-electric tensors satisfy:

$$\overline{\varepsilon}' = \varepsilon'_t(\mathbf{1}-\hat{\mathbf{x}}\otimes\hat{\mathbf{x}}) + \varepsilon'\hat{\mathbf{x}}\otimes\hat{\mathbf{x}}, \qquad\qquad \varepsilon'_t = \varepsilon'\frac{1-(v/c_0)^2}{1-(v/v_d)^2}, \qquad\qquad (5b)$$

$$\overline{\mu}' = \mu'_t(\mathbf{1}-\hat{\mathbf{x}}\otimes\hat{\mathbf{x}}) + \mu'\hat{\mathbf{x}}\otimes\hat{\mathbf{x}}, \qquad\qquad \mu'_t = \mu'\frac{1-(v/c_0)^2}{1-(v/v_d)^2}, \qquad\qquad (5c)$$

$$\overline{\xi}' = -\overline{\zeta}' = \frac{1-(c_0/v_d)^2}{1-(v/v_d)^2}\frac{c\mathbf{v}}{c_0^2}\times\mathbf{1}, \qquad\qquad (5d)$$

and $v_\mathrm{d} = c/\sqrt{\varepsilon'\mu'}$ is the velocity of the relevant dielectric in the laboratory frame. In the previous formulas $\varepsilon', \mu'$ stand for $\varepsilon' = \varepsilon(x'/\gamma, y', z')$ and $\mu' = \mu(x'/\gamma, y', z')$ with the functions in the right-hand side defined as in Eq. (1). Thereby, the material parameters are independent of time in the new coordinates. For a finite $c_0$, the $\gamma$-factor is greater than one, and hence all the lengths along the direction of motion in the lab frame are shorter than in the co-moving frame (the "rest" frame), due to the Lorentz-Fitzgerald length contraction [30]. Due to this reason, for a finite $c_0$ the geometry of the spacetime crystal in the laboratory frame is a contracted version of the geometry in the co-moving frame.

From Eq. (5) one sees that the constitutive relations in the co-moving frame are characterized by a bianisotropic coupling, described by the tensors $\overline{\overline{\xi'}} = -\overline{\overline{\zeta'}}$. As further discussed in the next subsection, this property is at odds with the response of moving isotropic dielectrics. Indeed, a moving dielectric has a response free of magneto-electric coupling in the rest frame (co-moving frame) and a bianisotropic response in any other inertial frame.

*B. Moving dielectric crystal*

It is relevant to contrast the response of a spacetime crystal with that of the corresponding moving photonic crystal. To do this, consider a time independent dielectric photonic crystal at rest in some inertial frame. In this frame (primed coordinates), the dielectric photonic crystal is characterized by standard constitutive relations:

$$\mathbf{D}'(\mathbf{r}',t') = \varepsilon_0 \varepsilon'(x', y', z') \mathbf{E}'(\mathbf{r}',t), \quad \mathbf{B}'(\mathbf{r}',t') = \varepsilon_0 \mu'(x', y', z') \mathbf{H}'(\mathbf{r}',t). \tag{6}$$

On the other hand, in a (laboratory) inertial frame that moves with speed $-v\hat{\mathbf{x}}$ with respect to the rest frame the transformed fields are related as [23] (here we use the standard Lorentz transformation with $c_0 = c$):

$$\begin{pmatrix} \mathbf{D} \\ \mathbf{B} \end{pmatrix} = \begin{pmatrix} \varepsilon_0 \bar{\varepsilon} & \dfrac{1}{c}\bar{\xi} \\ \dfrac{1}{c}\bar{\zeta} & \mu_0 \bar{\mu} \end{pmatrix} \begin{pmatrix} \mathbf{E} \\ \mathbf{H} \end{pmatrix} \qquad (7a)$$

where the relevant tensors are now given by:

$$\bar{\varepsilon} = \varepsilon_t \left(\mathbf{1} - \hat{\mathbf{x}} \otimes \hat{\mathbf{x}}\right) + \varepsilon \hat{\mathbf{x}} \otimes \hat{\mathbf{x}}, \qquad \varepsilon_t = \varepsilon \frac{1-(v/c)^2}{1-(v/v_d)^2}, \qquad (7b)$$

$$\bar{\mu} = \mu_t \left(\mathbf{1} - \hat{\mathbf{x}} \otimes \hat{\mathbf{x}}\right) + \mu \hat{\mathbf{x}} \otimes \hat{\mathbf{x}}, \qquad \mu_t = \mu \frac{1-(v/c)^2}{1-(v/v_d)^2}, \qquad (7c)$$

$$\bar{\xi} = -\bar{\zeta} = -\frac{1-(c/v_d)^2}{1-(v/v_d)^2} \frac{\mathbf{v}}{c} \times \mathbf{1}. \qquad (7d)$$

where $v_d = c/\sqrt{\varepsilon' \mu'}$. The parameters $\varepsilon, \mu$ are linked to $\varepsilon', \mu'$ as $\varepsilon = \varepsilon'(\gamma(x-vt), y, z)$ and $\mu = \mu'(\gamma(x-vt), y, z)$, so that the response in the laboratory frame is time dependent and has also a "travelling-wave" structure. Comparing Eqs. (1) and (7) and Eqs. (5) and (6), the difference between a moving dielectric crystal and spacetime modulated dielectric crystal becomes evident: the responses in the co-moving frame (where the constitutive relations are time independent in both problems) and in the laboratory frame (where the constitutive relations are time dependent) are swapped in the two problems. In particular, a moving photonic crystal is bianisotropic in the laboratory frame, while the corresponding spacetime crystal has a purely isotropic response in the laboratory frame. Clearly, the two types of crystals are generically rather different from an electromagnetic point of view.

*C. Isorefractive crystals and Minkowskian isotropic materials*

Let us now consider an isorefractive crystal such that $v_d$ is independent of space. In other words, the speed of light is identical in all the materials of the crystal. This type of crystals was discussed in Ref. [31] in a different context. Specifically, the authors analyzed the peculiar dispersion properties of light waves in such crystals near the transition between the subluminal and superluminal regimes. Furthermore, time independent isorefractive systems have been previously discussed in the literature in different contexts [32-34].

Consider first the ideal case $v_\mathrm{d} = c$, so that the speed of light in the materials is identical to the speed of light in vacuum. It should be noted that in realistic materials $v_\mathrm{d} < c$, as the light-matter interactions slow down the wave propagation with respect to the vacuum case. We will not worry with such a constraint for now; the requirement $v_\mathrm{d} = c$ will be relaxed below.

Using $v_\mathrm{d} = c$ and a standard Lorentz transformation ($c_0 = c$) in Eqs. (1), (5), (6) and (7), one readily finds that both for the spacetime crystal problem and for the moving photonic crystal problem the constitutive relations in the co-moving frame are of the type: $\mathbf{D}'(\mathbf{r}',t') = \varepsilon_0 \varepsilon'(x',y',z') \mathbf{E}'(\mathbf{r}',t)$ and $\mathbf{B}'(\mathbf{r}',t') = \varepsilon_0 \mu'(x',y',z') \mathbf{H}'(\mathbf{r}',t)$. Furthermore, when $v_\mathrm{d} = c = c_0$ the constitutive relations are preserved by the Lorentz transformation, i.e., they are frame independent. In a generic inertial frame, let us say the laboratory frame, they are of the form: $\mathbf{D} = \varepsilon_0 \varepsilon' \mathbf{E}$, $\mathbf{B} = \varepsilon_0 \mu' \mathbf{H}$. Note that in the laboratory frame the constitutive relations are time-dependent due to the moving material interfaces.

The enunciated results can be better understood noting that the standard Lorentz transformation preserves the constitutive relations of the electromagnetic vacuum, i.e., the vacuum is a "fixed point" of the Lorentz transformation. It has been previously noted [35] that there is a wider set of fixed points formed by all the isotropic "materials" with the same refractive index as the vacuum. Such class of materials is known as Minkowskian isotropic media.

The above discussion reveals that an arbitrary crystal formed by Minkowskian isotropic media ($v_\mathrm{d} = c$) is described by constitutive relations that are observer independent. Furthermore, it proves that a hypothetical moving photonic crystal formed by Minkowskian isotropic media has an electromagnetic response strictly equivalent to the response of the corresponding spacetime modulated crystal. Thereby, Minkowskian isotropic spacetime crystals may mimic perfectly the physical motion of some material body at *any* frequency. This is the first key result of the article.

As noted before, it is certainly challenging to implement Minkowskian spacetime crystals with $v_d = c$. However, one may relax the constraint $v_d = c$ so that it becomes $v_d = c_0$ where $c_0$ is now some arbitrary velocity, if desired much less than the speed of light in vacuum. Even though the response of such a spacetime crystal with $v_d = c_0$ is not strictly equivalent to that of a physical moving medium, in practice the two mathematical structures are rather similar. In fact, it should be obvious that an isorefractive spacetime crystal with $v_d = c_0$ effectively emulates a moving physical body in a fictitious "universe" where the speed of light is $c_0$, rather than $c$.

Indeed, from Eq. (5) the isorefractive materials characterized by a given velocity $v_d$ are "fixed points" of the generalized Lorentz transformation with $c_0 = v_d$. In other words, the generalized Lorentz transformation with $c_0 = v_d$ enables one to switch to a set of coordinates where the constitutive relations of the spacetime crystal remain precisely the same (i.e., described by a scalar permittivity and by a scalar permeability as in Eq. (1)), but time invariant. Thus, the electrodynamics of a generic isorefractive spacetime crystal is strictly determined by the electrodynamics of a standard time-independent photonic crystal through a generalized Lorentz transformation. This is the second key result of the article.

*D. Dispersion diagrams, generalized Doppler transformation, and addition of velocities*

From the previous subsection, the electrodynamics of isorefractive spacetime crystals can be conveniently studied in the co-moving frame where the constitutive relations of the material are isotropic and time-invariant. Note that this result holds true even for three-dimensional crystals.

Clearly, the electromagnetic modes in the co-moving frame coordinates are Bloch waves with a spacetime variation of the type: $e^{-i\omega' t'} e^{+i\mathbf{k}'\cdot\mathbf{r}'}$. The dispersion of the Bloch waves $\omega'$ vs. $\mathbf{k}'$ in the co-moving frame may be found with standard numerical methods. By applying an inverse generalized Lorentz transformation, one can obtain the modes in the original (unprimed) frame. The fields in the unprimed frame have a structure of the type $\mathbf{F}_\mathrm{p}(\mathbf{r}-\mathbf{v}t) e^{-i\omega t} e^{+i\mathbf{k}\cdot\mathbf{r}}$ with $\mathbf{F}_\mathrm{p}$ a periodic function in the three spatial coordinates. The dispersion in the laboratory frame is easily determined by a (generalized) Doppler transformation [30]:

$$k_x = \gamma\left(k_x' + v\frac{\omega'}{c_0^2}\right), \qquad k_y = k_y', \qquad k_z = k_z'. \tag{8a}$$

$$\omega = \gamma\left(\omega' + k_x' v\right). \tag{8b}$$

It is interesting to relate the wave velocities in the co-moving and laboratory frames. For simplicity, we restrict our discussion to the case of Bloch modes that propagate along the $x$-direction (i.e., along the direction parallel to $\mathbf{v}$) in the long wavelength limit. Clearly, as in the co-moving frame the system is a conventional reciprocal photonic crystal, the velocities of the waves that propagate along the $+x$ and $-x$ directions differ by a minus sign: $v_w'^+ = -v_w'^-$. The velocities are determined by the slopes $v_w'^\pm = \omega'/k_x'^\pm$ evaluated in the long wavelength limit.

The wave velocities in the laboratory frame ($v_w^\pm = \omega/k_x^\pm$) can be readily determined using the (generalized) formula for the relativistic addition of velocities [30]:

$$v_w^\pm = \frac{v_w'^\pm + v}{1 + v_w'^\pm v/c_0^2}. \tag{9}$$

Clearly, $|v_w^+| > |v_w^-|$ in the subluminal regime ($|v| < c_0$ and $|v_w'| < c_0$) and thereby the synthetic motion always induces a "positive" Fresnel drag effect, such that the waves in the laboratory frame propagate faster along the direction determined by the synthetic motion (+x-direction). It is underscored that different from the general case considered in previous works [19, 20], here the Fresnel drag effect is determined by a relativistic addition formula, due to the rigorous analogy between the isorefractive spacetime crystal and a moving crystal.

*E. Numerical examples*

In order to illustrate the ideas of the previous subsections, next we present two numerical examples. In the first example, the spacetime crystal in the co-moving frame is formed by an isorefractive honeycomb array of dielectric cylinders with radius $R'$ [Fig. 1a]. For simplicity, we consider the case of Minkowskian isotropic crystals, so that the background region is air and the cylinders have permittivity $\varepsilon_A'$ and permeability $\mu_A'$ constrained by $n_A' = \sqrt{\varepsilon_A' \mu_A'} = 1$ (in the numerical simulations we take $\varepsilon_A' = 5$, $\mu_A' = 1/\varepsilon_A'$). The distance between nearest neighbors is $a'$.

We consider waves with transverse electric (TE) polarization ($\mathbf{E}' = E_z' \hat{\mathbf{z}}'$) and propagation in the *xoy* plane, so that $E_z' = E_z'(x', y')$. The band structure in the co-moving frame can be found by solving the secular equation:

$$-\partial_{x'}\left(\frac{1}{\mu'(x',y')}\partial_{x'}E_{z'}'\right) - \partial_{y'}\left(\frac{1}{\mu'(x',y')}\partial_{y'}E_{z'}'\right) = \left(\frac{\omega'}{c}\right)^2 \varepsilon'(x',y') \, E_{z'}', \qquad (10)$$

where $x', y', z'$ are the spatial coordinates in the co-moving frame. The Bloch modes are calculated using the plane wave method [36].

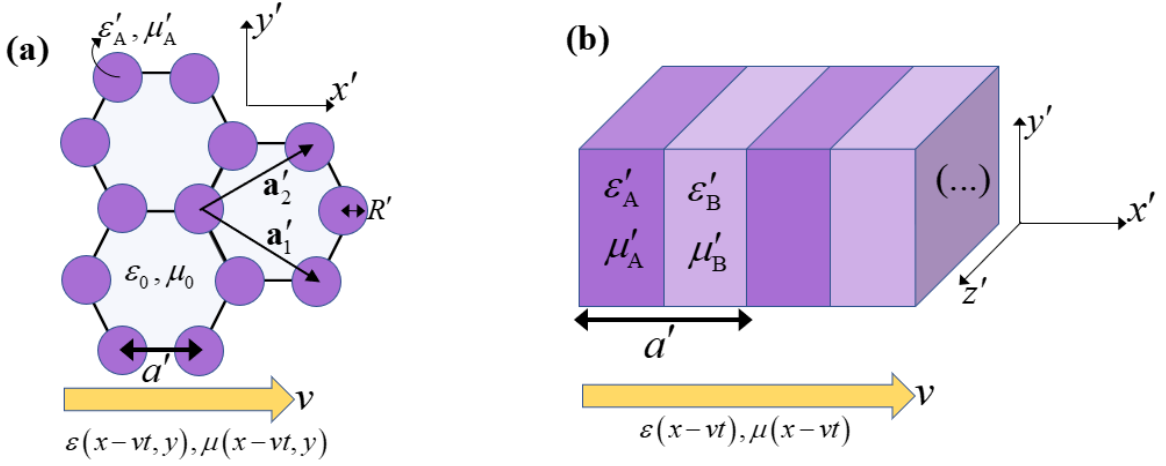

FIG. 1. Geometry of two isorefractive spacetime crystals with a travelling-wave modulation. The modulation speed is $\mathbf{v} = v\hat{\mathbf{x}}$. The arrow indicates how the material parameters vary in time in the laboratory frame. **(a)** Co-moving frame geometry of a honeycomb array of dielectric scatterers with radius $R'$ and permittivity and permeability $\varepsilon'_A, \mu'_A$ embedded in air, with $n'_A = \sqrt{\varepsilon'_A \mu'_A} = 1$. The direct lattice primitive vectors are $\mathbf{a}'_1 = a'/2(3, -\sqrt{3})$ and $\mathbf{a}'_2 = a'/2(3, \sqrt{3})$, where $a'$ is the distance between nearest neighbors. **(b)** Stratified isorefractive spacetime crystal formed by two isotropic layers with material parameters $\varepsilon'_A, \mu'_A$ and $\varepsilon'_B, \mu'_B$, such that $\sqrt{\varepsilon'_A \mu'_A} = 1 = \sqrt{\varepsilon'_B \mu'_B}$. The lattice period in the co-moving frame is $a'$.

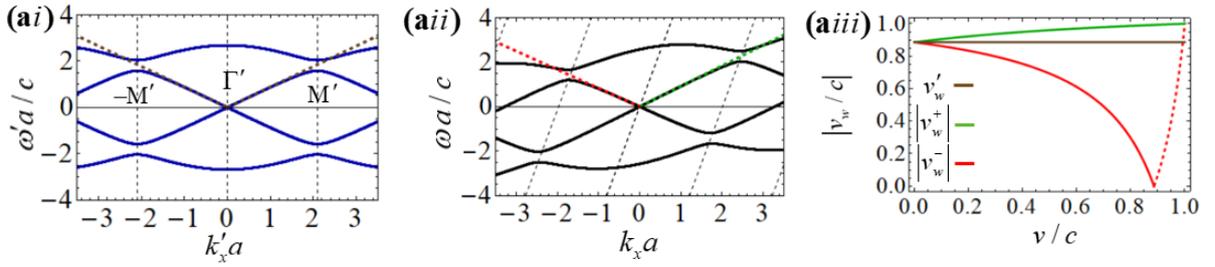

FIG. 2. (a) Exact dispersion diagram of the isorefractive spacetime honeycomb crystal [Fig. 1a] for $\varepsilon'_A = 5$, $\mu'_A = 1/\varepsilon'_A$ for a modulation speed $v = 0.2c$ calculated in the (a*i*) co-moving frame, and (a*ii*) laboratory frame. (a*iii*) Long wavelength limit wave velocities in the laboratory frame $v_w^+, v_w^-$ as a function of the modulation speed $v/c$. The horizontal line represents the wave velocity in the co-moving frame $|v'_w| = |v'^\pm_w|$. In the dashed part of the red curve, $v_w^-$ is positive and the propagation is unidirectional.

The numerically calculated band diagram is plotted in Fig. 2*ai* for the parameters $\varepsilon'_A = 5$, $\mu'_A = 1/\varepsilon'_A$, $R' = 0.4a$ and a modulation speed $v = 0.2c$. It should be noted that due to the Lorentz-Fitzgerald contraction the cross-section of the cylinders in the laboratory frame coordinates is ellipsoidal rather than circular. Furthermore, the original honeycomb lattice is slightly contracted for the same reason. For simplicity, we restrict our attention to the $x'$-axis and to the segment of the Brillouin zone $-M' \to \Gamma' \to M'$ (corresponding to the vertical dashed lines in Fig. 2ai). Here, $\Gamma', M'$ are the standard high-symmetry points of the honeycomb array. In the long wavelength limit, the photonic crystal dispersion can be approximated by two lines depicted in Fig. 2ai with the dashed black style, whose slopes determine the wave velocities $v'^{\pm}_w = \omega'/k'^{\pm}_x$ along the $\pm x'$-axis. As already mentioned, due to Lorentz reciprocity the two slopes are identical in the co-moving frame: $|v'^{+}_w| = |v'^{-}_w| \equiv v'_w$.

The band diagram in the laboratory frame [Fig. 2aii] is found with the help of the relativistic Doppler transformation [Eq. (8)]. Due to the Doppler shift, the band structure in the laboratory frame is tilted with respect to the co-moving frame. The synthetic motion creates an evident spectral asymmetry, $\omega(k_x) \neq \omega(-k_x)$. In particular, there is a synthetic Fresnel drag such that the wave velocities in the long wavelength limit obey the order relation: $|v^{-}_w| < v < |v^{+}_w|$.

Figure 2aiii depicts the velocities in the laboratory frame $|v^{\pm}_w|$ as a function of the modulation speed. The plot is obtained using the relativistic addition of velocities [Eq. (9)]. Note that the crystal geometry in the co-moving frame is assumed independent of the modulation speed. As seen, the asymmetry between the velocities of counter-propagating waves becomes more pronounced as the modulation speed increases. Interestingly, for $v = v'_w = |v'^{\pm}_w|$, the velocity of the counter-propagating wave ($v^{-}_w$) becomes exactly zero.

Furthermore, for $v > v'_w$ the signs of the velocities of the two waves $v^{\pm}_w$ become identical. Thus, the velocity $v = v'_w$ marks the transition between the usual bi-directional propagation regime ($v^+_w$ and $v^-_w$ have opposite signs) and a unidirectional propagation regime ($v^+_w$ and $v^-_w$ have both positive signs), in agreement with the findings of Ref. [31]. As the modulation speed approaches the superluminal threshold ($v \to c$), both $v^{\pm}_w$ approach $+c$.

As a second example, we consider a 1D-type photonic crystal formed by a periodic stack of isorefractive dielectric layers with period $a'$ in the co-moving frame [Fig. 1b]. The material parameters are taken as $\varepsilon'_A = 2$, $\mu'_A = 1/2$ and $\varepsilon'_B = 1/4$, $\mu'_B = 4$, such that the refractive indexes, are $n'_A = n'_B = 1$. The band diagrams calculated in the co-moving and laboratory frames are shown in Figs. 3ai and 3aii, respectively, for the modulation velocities $v = 0.1c$ and $v = 0.35c$. Similar to the previous example, the spectral symmetry is broken in the laboratory frame ($\omega(k_x) \neq \omega(-k_x)$) due to the synthetic motion.

It is relevant to discuss the homogenization and long wavelength limit response of the Minkowskian spacetime crystal [18, 20, 21]. As is well-known, for stratified structures, the effective response in the co-moving frame can be found with simple spatial averaging of the material parameters. For layers with identical thickness the effective permittivity and permeability are:

$$\varepsilon'_{\text{ef,L}} = \frac{\varepsilon'_A + \varepsilon'_B}{2}, \qquad \mu'_{\text{ef,L}} = \frac{\mu'_A + \mu'_B}{2}. \tag{11}$$

The effective parameters describe the response of the crystal to transverse waves that propagate along the direction of motion. The important point is that even though the two material layers are isorefractive ($n'_A = n'_B = 1$), the effective medium has typically a different refractive index: $\sqrt{\varepsilon'_{\text{ef,L}} \mu'_{\text{ef,L}}} \neq 1$. This implies that the effective medium is not a "fixed point" of the Lorentz transformation, different from the materials A and B. In fact, the effective material parameters in the lab frame can be readily found with the help of Eq. (7) using the substitution $v_d \to v_{d,ef} = \frac{c}{\sqrt{\varepsilon'_{ef,L} \mu'_{ef,L}}}$. For waves polarized in the $yoz$ plane the effective response is determined by:

$$\varepsilon_{ef} = \varepsilon'_{ef,L} \frac{1-(v/c_0)^2}{1-(v/v_{d,ef})^2}, \quad \mu_{ef} = \mu'_{ef,L} \frac{1-(v/c_0)^2}{1-(v/v_{d,ef})^2}, \quad \xi_{ef} = \frac{1-(c_0/v_{d,ef})^2}{1-(v/v_{d,ef})^2} \frac{cv}{c_0^2}, \quad (12)$$

with $c_0 = v_d$. The parameter $\xi_{ef}$ determines the effective magneto-electric tensor $\overline{\overline{\xi}} = -\xi_{ef} \hat{\mathbf{x}} \times \mathbf{1}$ in the laboratory frame. In the previous discussion, it is implicit that $c_0 = c$, but the above formula remains valid for an arbitrary value of $c_0 = v_d$.

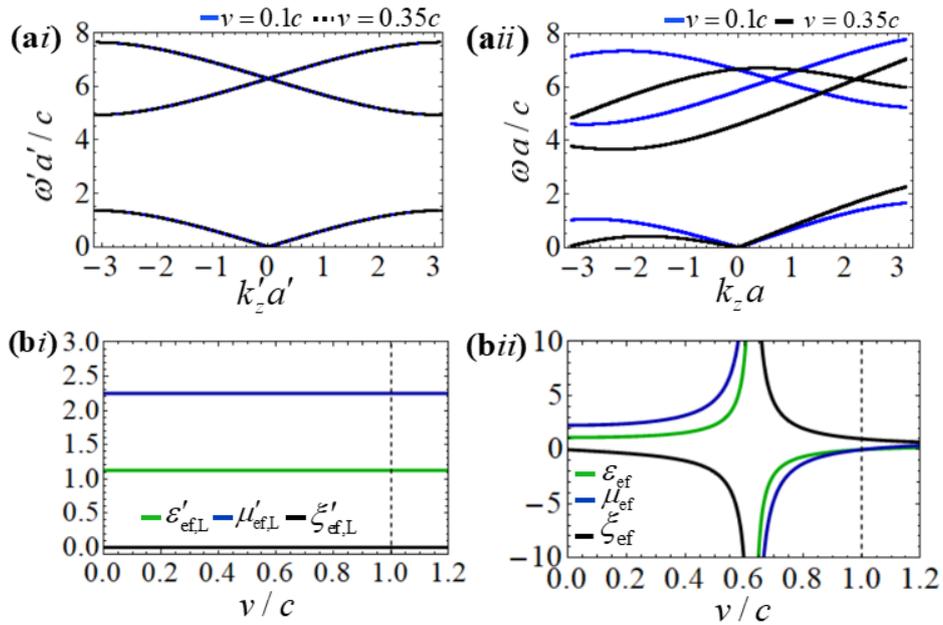

FIG. 3. (a) Exact dispersion diagrams of the isorefractive spacetime crystal formed by two dielectric layers [Fig. 1b] with parameters $\varepsilon'_A = 2$, $\mu'_A = 1/2$ and $\varepsilon'_B = 1/4$, $\mu'_B = 4$, for the modulation speeds $v = 0.1c$ and $v = 0.35c$. (ai) Co-moving frame diagrams, (aii) Laboratory frame diagrams. (b) Effective parameters as a function of the modulation speed. (bi) $\varepsilon'_{ef,L}, \mu'_{ef,L}, \xi'_{ef,L}$ in the co-moving frame. (bii) $\varepsilon_{ef}, \mu_{ef}, \xi_{ef}$ in the laboratory frame. Note that the response in the co-moving frame is isotropic ($\xi'_{ef,L} = 0$). The vertical dashed line marks the transition between the subluminal and superluminal regimes.

Figures 3bi) and 3bii) depict the effective parameters of the 1D spacetime crystal in the co-moving and laboratory frames as a function of the modulation speed. Again, it is assumed that the geometry of the photonic crystal in the co-moving frame is independent of $v$. As seen in Fig. 3bii), the effective parameters in the laboratory frame depend on the modulation speed $v$ due to $\sqrt{\varepsilon'_{\text{ef,L}} \mu'_{\text{ef,L}}} \neq 1$. In particular, the magneto-electric parameter $\xi_{\text{ef}}$ diverges and switches sign for $v = v_{\text{d,ef}} < c$, consistent with Eq. (12). The effective permittivity and permeability $\varepsilon_{\text{ef}}, \mu_{\text{ef}}$ exhibit a similar behavior, so that all the effective parameters are resonant within the subluminal range. This resonance marks the transition between the bi-directional and the unidirectional propagation regimes, already discussed in the first example.

It is important to underline that the previous discussion can be readily generalized to the homogenization of 2D and 3D isorefractive spacetime crystals. In fact, for any effective medium model developed in the co-moving frame (e.g., relying on standard mixing formulas such as the Maxwell-Garnett formula, or others), the corresponding effective parameters in the laboratory frame can be readily determined with the help of Eq. (12).

### III.  Comparison of Galilean and Lorentz transformations

In recent works, the response of spacetime crystals was studied with a Galilean transformation of coordinates, such that $\mathbf{r}' = \mathbf{r} - \mathbf{v}t$ and $t' = t$, e.g., [20, 21]. The key property of the Galilean transformation is that it preserves the structure of the Maxwell's equations so that in the co-moving frame coordinates one has $\nabla' \times \mathbf{E}' = -\partial_{t'}\mathbf{B}'$, $\nabla' \times \mathbf{H}' = +\partial_{t'}\mathbf{D}'$, similar to the Maxwell's equations in the laboratory frame $\nabla \times \mathbf{E} = -\partial_t \mathbf{B}$, $\nabla \times \mathbf{H} = +\partial_t \mathbf{D}$. The fields in the co-moving frame are linked by bianisotropic constitutive relations that are time-independent [Eq. (5) with $c_0 = \infty$].

The Galilean transformation is rather convenient from a computational standpoint as it does not mix the time and space coordinates. In particular, the geometry of the problem is identical in the laboratory and co-moving frames due to the absence of the Lorentz-Fitzgerald length contraction. Furthermore, the Galilean transformation is particularly useful in the superluminal range, where the Lorentz transformation breaks down and the $\gamma$-factor becomes purely imaginary. Evidently, the fields associated with the Galilean transformation [defined by Eq. (3) with $c_0 = \infty$] are deprived of an have an immediate physical meaning, and should be simply regarded as auxiliary fields that are introduced to find the physical fields in the laboratory frame.

More generally, it is possible to study the electrodynamics of a travelling-wave spacetime crystal using any of the Lorentz transformations defined by Eq. (2). While the fields, the constitutive relations, the band diagrams, etc, in the co-moving frame typically depend on the considered $c_0$, the corresponding quantities in the laboratory frame are independent of $c_0$, if the inverse transformation is correctly applied. The fields in the co-moving frame coincide with the physical fields in the corresponding inertial frame only when $c_0 = c$. A relativistic Lorentz transformation with $c_0 = v_d$ is particularly useful in the case of isorefractive (Minkowskian) spacetime crystals as it leads to simple observer-independent isotropic constitutive relations, different from the Galilean transformation which leads to a bianisotropic response.

It is less obvious if the spacetime crystal homogenization via a Galilean transformation, as presented in Refs. [20, 21], necessarily agrees with the homogenization achieved through a relativistic transformation. The objective of the rest of this section is to show that indeed the two types of transformations yield identical effective parameters in the laboratory frame. The following analysis is not restricted to isorefractive materials.

The homogenization methodology follows the same steps as in Sect. II.E (see Refs. [20, 21] for more details). First, using a Galilean or a Lorentz transformation we switch to a co-moving frame where the material parameters are independent of time. For stratified crystals the effective parameters in the co-moving frame are determined by the spatial average of the co-moving frame parameters [20, 21]. Finally, the response in the laboratory frame is determined using an inverse Galilean or Lorentz transformation. Different from section II, in the following the geometry of the crystal is fixed in the laboratory frame, rather than in the Lorentz co-moving frame. Thus, the thickness of the material layers is now fixed in the laboratory frame.

Applying the outlined procedure to a bi-layer crystal formed by dielectric slabs A and B with the same thickness [Fig. 1b], one finds that with a Lorentz transformation (Eq. (5) with $c_0 = c$) the effective parameters in the co-moving frame are:

$$\varepsilon'_{\text{ef,L}} = \left(1 - \frac{v^2}{c^2}\right)\frac{1}{2}\left(\frac{\varepsilon_{\text{A}}}{1 - n_{\text{A}}^2 \frac{v^2}{c^2}} + \frac{\varepsilon_{\text{B}}}{1 - n_{\text{B}}^2 \frac{v^2}{c^2}}\right), \quad \mu'_{\text{ef,L}} = \left(1 - \frac{v^2}{c^2}\right)\frac{1}{2}\left(\frac{\mu_{\text{A}}}{1 - n_{\text{A}}^2 \frac{v^2}{c^2}} + \frac{\mu_{\text{B}}}{1 - n_{\text{B}}^2 \frac{v^2}{c^2}}\right) \quad (13a)$$

$$\xi'_{\text{ef,L}} = \frac{1}{2}\frac{v}{c}\left(\frac{n_{\text{A}}^2 - 1}{1 - n_{\text{A}}^2 \frac{v^2}{c^2}} + \frac{n_{\text{B}}^2 - 1}{1 - n_{\text{B}}^2 \frac{v^2}{c^2}}\right), \quad (13b)$$

where $\varepsilon'_{\text{ef,L}}, \mu'_{\text{ef,L}}$ are the effective permittivity and permeability in the Lorentz co-moving frame, respectively, $\xi'_{\text{ef,L}}$ is the moving medium parameter such that $\overline{\xi'_{\text{ef}}} = -\xi'_{\text{ef,L}}\hat{\mathbf{x}} \times \mathbf{1}$, $n_i = \sqrt{\varepsilon_i \mu_i}$ ($i$=A, B) are the refractive indices of the layers A and B. Remarkably, the effective parameters in Eq. (13) do not depend on the Lorentz factor $\gamma$. When $n_{\text{A}} = n_{\text{B}} = 1$ the above formula reduces to Eq. (11) and the response becomes purely isotropic in the co-moving frame ($\xi'_{\text{ef,L}} = 0$).

On the other hand, using a Galilean transformation (Eq. (5) with $c_0 = \infty$) one finds that the corresponding effective parameters are:

$$\varepsilon'_{ef,G} = \frac{1}{2}\left(\frac{\varepsilon_A}{1-n_A^2\frac{v^2}{c^2}} + \frac{\varepsilon_B}{1-n_B^2\frac{v^2}{c^2}}\right), \quad \mu'_{ef,G} = \frac{1}{2}\left(\frac{\mu_A}{1-n_A^2\frac{v^2}{c^2}} + \frac{\mu_B}{1-n_B^2\frac{v^2}{c^2}}\right), \quad (14a)$$

$$\xi'_{ef,G} = \frac{1}{2}\frac{v}{c}\left(\frac{n_A^2}{1-n_A^2\frac{v^2}{c^2}} + \frac{n_B^2}{1-n_B^2\frac{v^2}{c^2}}\right). \quad (14b)$$

Clearly, the effective parameters in the two co-moving frames are different. Curiously, the effective permittivity and permeability $\varepsilon'_{ef,L}, \mu'_{ef,L}$ in the Lorentz co-moving frame can be written in terms of the parameters in the Galilean co-moving frame as $\varepsilon'_{ef,L} = \left(1-\frac{v^2}{c^2}\right)\varepsilon'_{ef,G}$ and $\mu'_{ef,L} = \left(1-\frac{v^2}{c^2}\right)\mu'_{ef,G}$. It is relevant to note that in the Galilean framework the magneto-electric coupling parameter $\xi'_{ef,G}$ does not vanish for isorefractive crystals.

Using either Eq. (13) combined with $c_0 = c$ or Eq. (14) combined with $c_0 = \infty$, one can calculate the effective parameters in laboratory frame with Eqs. (3b) and (4b). Importantly, it turns out that the effective parameters in the laboratory frame are independent of $c_0$, i.e. are independent if one uses a Lorentz or a Galilean transformation. They can be written explicitly as:

$$\varepsilon_{ef} = \frac{\varepsilon_A + \varepsilon_B - \varepsilon_A\varepsilon_B(\mu_A + \mu_B)\frac{v^2}{c^2}}{2 - \frac{1}{2}(\varepsilon_A + \varepsilon_B)(\mu_A + \mu_B)\frac{v^2}{c^2}}, \quad \mu_{ef} = \frac{\mu_A + \mu_B - \mu_A\mu_B(\varepsilon_A + \varepsilon_B)\frac{v^2}{c^2}}{2 - \frac{1}{2}(\varepsilon_A + \varepsilon_B)(\mu_A + \mu_B)\frac{v^2}{c^2}}, \quad (15a)$$

$$\xi_{ef} = \frac{v}{2c}\frac{(\varepsilon_A - \varepsilon_B)(\mu_A - \mu_B)}{2 - \frac{1}{2}(\varepsilon_A + \varepsilon_B)(\mu_A + \mu_B)\frac{v^2}{c^2}}. \quad (15b)$$

For an isorefractive system, $v_d = 1/\sqrt{\varepsilon_A \mu_A} = 1/\sqrt{\varepsilon_B \mu_B}$, the above formulas reduce to Eq. (12).

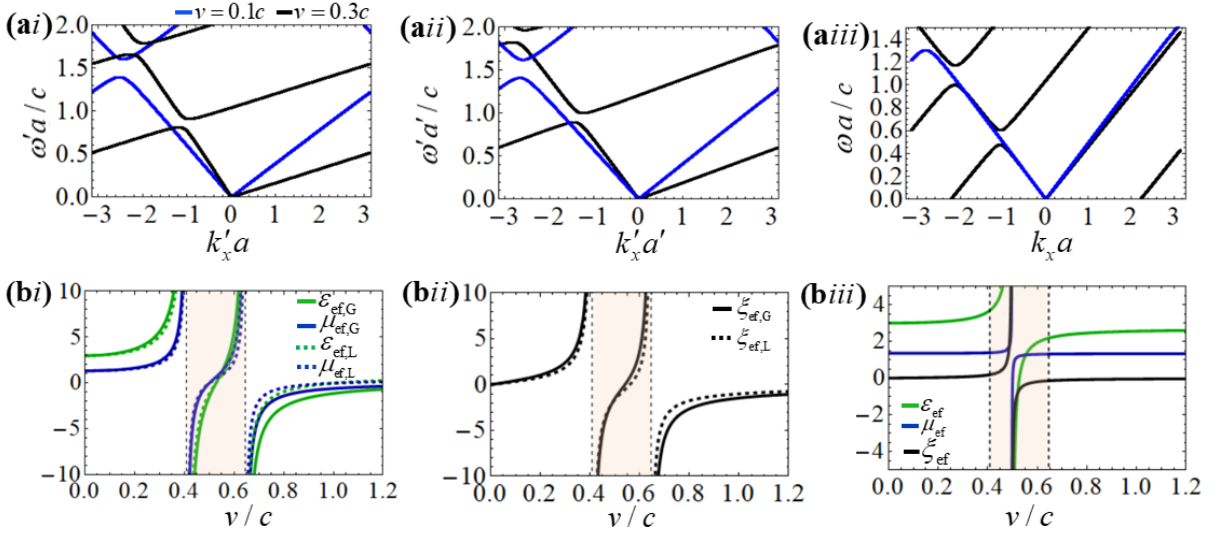

FIG. 4. **(a)** Dispersion diagrams of a spacetime crystal [Fig. 1b] formed by two isotropic dielectric layers with parameters $\varepsilon_A = 4$, $\mu_A = 1.5$ and $\varepsilon_B = 2, \mu_B = 1.2$. The dispersion diagrams are calculated for the modulation speeds are $v = 0.1c$ (blue lines) and $v = 0.3c$ (black lines) in the **(a*i*)** Galilean co-moving frame, **(a*ii*)** Lorentz co-moving frame, and **(a*iii*)** in the laboratory frame. **(b)** Effective parameters of the spacetime crystal as a function of the modulation speed. **(b*i*)** permittivity and permeability in the Galilean and Lorentz co-moving frames, **(b*ii*)** magneto-electric coupling parameter in the Galilean and Lorentz co-moving frames, **(b*iii*)** effective parameters in the laboratory frame. The shaded region in the plots represents the transluminal region ($c/n_A < v < c/n_B$) that separates the subluminal and superluminal regimes.

To illustrate the discussion, we represent in Fig. 4a, the dispersion diagram of a bi-layer spacetime crystal in the laboratory frame and in the Galilean and Lorentz co-moving frames for the modulation speeds $v = 0.1c$ and $v = 0.3c$. The dispersion diagram is calculated using the formalism of the Appendix. As seen, the dispersion diagrams in the co-moving Lorentz and Galilean frames do not coincide. However, when the inverse Doppler shift is applied to diagrams one obtains a consistent result, so that the dispersion in the laboratory frame is independent of the transformation, as it should be.

Figure 4b represents the effective parameters of the same spacetime crystal calculated using Eqs. (13) and (14) as a function of the modulation speed. For large modulation speeds, there is an evident difference between the parameters of the Galilean and Lorentz co-moving frames (see Figs. 4bi and 4bii). In both co-moving frames, the effective parameters diverge at the luminal transitions $v = c/n_A$ and $v = c/n_B$. In contrast, in the laboratory frame the effective parameters diverge inside the transluminal region ($c/n_A < v < c/n_B$).

## IV. Conclusion

In summary, we introduced the concept of Minkowskian isorefractive spacetime crystals, as time-variant systems described by constitutive relations that are observer independent. It was shown that ideal Minkowskian crystals with $v_d = c$ can replicate exactly the response of a moving dielectric photonic crystal for any frequency of operation. In particular, the band diagram of a Minkowskian crystal may be calculated with standard numerical methods and a relativistic Doppler transformation. In addition, the synthetic Fresnel drag can be rigorously characterized with the relativistic addition formula. More generally, it was shown that the more practical class of isorefractive spacetime crystals with $v_d = c_0 < c$ has a mathematical structure rather similar to that of Minkowskian crystals. Thereby, such systems can be analyzed and studied using essentially the same mathematical tools. We applied the theory to one-dimensional and two-dimensional isorefractive crystals, showing that it greatly simplifies the analysis and the understanding of the physical response of such systems. Furthermore, we discussed the impact of using relativistic and non-relativistic transformation of coordinates in the analysis of travelling-wave spacetime crystals. It was highlighted that relativistic and non-relativistic transformations predict exactly the same results for the band diagrams and effective response in the laboratory frame. We believe that isorefractive spacetime crystals provide an ideal platform to mimic physical motion across a wide frequency range.

**Acknowledgements:** This work is supported in part by the IET under the A F Harvey Engineering Research Prize, by the Simons Foundation, and by Fundação para a Ciência e a Tecnologia and Instituto de Telecomunicações under project UID/EEA/50008/2020.

## Appendix: Band structure in the Lorentz and Galilean co-moving frames

In this Appendix, we briefly explain how to calculate the band structure of a generic stratified spacetime crystal in the Galilean and Lorentz co-moving frames. We consider transverse electromagnetic waves propagating along the $x'$-direction. Following Ref. [21], the Maxwell's equations in the Galilean or Lorentz co-moving frames can be rewritten in a 4×4 matrix form as:

$$\frac{d}{dx'}\begin{pmatrix} \mathbf{E}'_\perp \\ \mathbf{H}'_\perp \end{pmatrix} = -i\omega' \boldsymbol{\sigma} \cdot \mathbf{M}'_\perp \cdot \begin{pmatrix} \mathbf{E}'_\perp \\ \mathbf{H}'_\perp \end{pmatrix} \tag{A1}$$

with

$$\boldsymbol{\sigma} = \begin{pmatrix} \mathbf{0}_{2\times 2} & \mathbf{J} \\ -\mathbf{J} & \mathbf{0}_{2\times 2} \end{pmatrix}, \quad \text{and} \quad \mathbf{J} = \begin{pmatrix} 0 & -1 \\ 1 & 0 \end{pmatrix}. \tag{A2}$$

Here, $\mathbf{E}'_\perp = \begin{pmatrix} E'_y & E'_z \end{pmatrix}^T$ and $\mathbf{H}'_\perp = \begin{pmatrix} H'_y & H'_z \end{pmatrix}^T$ are the transverse fields in the relevant co-moving frame. The primed material matrix in the co-moving frame $\mathbf{M}'_\perp$ is defined by [21]

$$\mathbf{M}'_\perp = \left[\frac{1}{c_0^2} v\boldsymbol{\sigma} + \mathbf{M}_\perp\right] \cdot \left[\mathbf{1}_{4\times 4} + v\boldsymbol{\sigma} \cdot \mathbf{M}_\perp\right]^{-1}, \tag{A3a}$$

where $c_0 = c$ or $c_0 = \infty$ for the Lorentz and Galilean cases, respectively. In the above, $\mathbf{M}_\perp$ is the transverse material matrix in the laboratory frame, defined in terms of the permittivity and permeability [21]:

$$\mathbf{M}_\perp = \begin{pmatrix} \varepsilon_0 \varepsilon \mathbf{1}_{2\times 2} & \mathbf{0}_{2\times 2} \\ \mathbf{0}_{2\times 2} & \mu_0 \mu \mathbf{1}_{2\times 2} \end{pmatrix}. \tag{A3b}$$

Following Ref. [21], for a two-phase crystal with layers *A* and *B* of identical thickness (half-lattice constant, $a'/2$) the dispersion $\omega'$ vs. $k'_x$ in the co-moving frame is determined by

$$\det\left(\exp\left(-i\frac{\omega' a'}{2}\boldsymbol{\sigma}\cdot\mathbf{M}'_{\perp,B}\right)\cdot\exp\left(-i\frac{\omega' a'}{2}\boldsymbol{\sigma}\cdot\mathbf{M}'_{\perp,A}\right)-e^{ik'_x a'}\mathbf{1}_{4\times 4}\right)=0, \tag{A4}$$

where $\mathbf{M}'_{\perp,i}$ is the (transverse) material matrix for layer $i$=A, B and $\exp(...)$ stands for the exponential of a matrix. Note that the lattice constant in the co-moving frame is related to the lattice constant in the laboratory frame as $a'=\gamma a$. The dispersion $\omega$ vs. $k_x$ in the laboratory frame can be found using the Doppler transformation (8).